\documentstyle[preprint,aps,floats,tighten]{revtex}
\begin{document}
\draft
\preprint{\vbox{\hbox{CU-TP-754}
                \hbox{CAL-607}
                \hbox{FERMILAB--Pub--96/123-A}
                \hbox{astro-ph/9606132}
}}

\title{Comment on ``The Dispersion Velocity of Galactic Dark
Matter Particles''}

\author{Evalyn
Gates$^{1,2}$\footnote{gates@tyrone.uchicago.edu}, Marc
Kamionkowski$^4$\footnote{kamion@phys.columbia.edu}, and Michael
S. Turner$^{1,2,3}$\footnote{mturner@fnal.fnal.gov}}
\address{$^1$NASA/Fermilab Theoretical Astrophysics Center,
Fermi National Accelerator Laboratory, Batavia, IL 60615-0500}
\address{$^2$Department of Astronomy \& Astrophyisics, Enrico
Fermi Institute, University of Chicago, Chicago, IL 60637-1433}
\address{$^3$Department of Physics, Enrico Fermi Institute,
University of Chicago, Chicago, IL 60637-1433}
\address{$^4$Department of Physics, Columbia University, 538 West
120th St., New York, New York~~10027}
\date{June 1996}
\maketitle



\vskip 1cm

In a recent Letter \cite{cowsik}, Cowsik et al. claim that a
self-consistent treatment of the dark Galactic halo, which
takes into account the gravitational effect of
luminous matter and allows for a nonspherical
halo, requires that the local velocity dispersion of
dark-matter particles be $600$ km~s$^{-1}$ or greater,
more than a factor of two
larger than the canonical value of 270 km~s$^{-1}$.  If true,
this would significantly affect rates and signatures for
detection of baryonic and nonbaryonic dark matter.

This work contradicts the assembled
results of a long history of work in Galactic dynamics,
which among other things holds that the velocity dispersion of the
halo should be close to 270 km s$^{-1}$, the value that obtains for
a spherically symmetric isothermal halo, $\sqrt{3/2}$ times
the asymptotic rotation velocity of around 220\,km\,s$^{-1}$.
We believe that this work
is incorrect, probably because not all the observational constraints
were taken into account and because the models were forced to satisfy
arbitrary constraints on the halo density.

Cowsik et al. construct their models for the distribution of
halo dark-matter particles by assuming an isothermal (i.e.,
Maxwell-Boltzmann velocity distribution with a constant
dispersion), axisymmetric
distribution of dark-matter particles that move in
the combined gravitational potential of the bulge, disk, and
halo.  They solve the coupled Boltzmann and Newton equations
iteratively, subject to the arbitrary boundary conditions:
$\rho_{\rm DM}(r=0) \sim 1\,{\rm GeV}/{\rm cm}^3$ and $\rho_{\rm DM}
(r=8\,{\rm kpc}) \sim 0.3\,{\rm GeV}/{\rm cm}^3$.  We call these
arbitrary because the density of dark-matter particles at the center
of the Galaxy and at the solar circle are not observed quantities,
but are derived from models for the Galaxy.
They obtain the velocity dispersion of 600 km\,s$^{-1}$ (or
larger for other assumed values of the local projected mass
density of the disk) by fitting the calculated
equatorial rotation curve for the model to the data from 2\,kpc
to 18\,kpc.

There are a number of things we do not know about their results,
e.g., how broad is the minimum of their $\chi^2$ and does it
include the canonical value for the halo velocity dispersion?
Why is it that the curves in their Figure do not asymptote to 
$\sqrt{2/3}$ times the velocity dispersion as they should for an
isothermal halo?

However, we do know that their model conflicts
with several important observational facts.  In the neighborhood of the
solar circle the velocity dispersion of the halo has been determined
{}from velocity measurements of halo stars and globular clusters and is
found to be around 200\,km\,s$^{-1}$ \cite{wyseannrev}.  Since
these objects trace the halo gravitational potential this is in
severe conflict with their result.  
In addition, the rotation curves for
several of their preferred halo models exceed 250\,km\,s$^{-1}$
at 20\,kpc and are still rising.  This is likely to be in conflict
with determinations of the rotation speed ($\simeq$200--250 km
s$^{-1}$) at distances from 30\,kpc to
100\,kpc based upon the proper motions of the Milky Way's satellites.

Finally, others have studied the effect of the bulge and disk on the
halo as well as flattening of the halo and find that the velocity dispersion
in the halo is not changed significantly.  That the bulge and
disk do not affect the halo is easily understood:  the mass
of the bulge is small ($\sim 2\times 10^{10}M_\odot$) and so its
effects are restricted to near the center of the Galaxy; the velocity
dispersion within the disk is only around 30\,km\,s$^{-1}$.
While flattening the halo can increase the local halo density
\cite{ggt}, it can be shown by use of the virial theorem that
it does not significantly affect the velocity dispersion.
Kuijken and Dubinski \cite{kuijken} find that the local halo
velocity dispersions in several self-consistent  models for the
disk, bulge, and halo of the Milky Way range from 246 to 323
\,km\,s$^{-1}$.

Without repeating their numerical calculations we cannot be certain where
Cowsik et al. went wrong.  However, we are confident that their
lower limit to the dark-matter velocity dispersion is not
correct because it disagrees with observations and because
previous work found that the effect of the luminous matter on
the halo was small.

\end{document}